\documentclass[14pt, oneside]{article}

\usepackage[letterpaper, margin=1in, includeheadfoot]{geometry}
\usepackage{hyperref}
\usepackage{tabularx}
\usepackage{subcaption}
\usepackage{fancyhdr}
\usepackage{graphicx}
\usepackage{amssymb}
\usepackage{amsmath}
\usepackage[numbers,square,sort]{natbib}

\begin{document}


\begin{centering}
{\Large \textbf{A Multi-loudspeaker Binaural Room Impulse Response Dataset with High-Resolution Translational and Rotational Head Coordinates in a Listening Room}}\\
\vspace{\baselineskip}
Yue Qiao$^{*}$, Ryan Miguel Gonzales, and Edgar Choueiri\\
\vspace{\baselineskip}
Correspondence$^{*}$: Yue Qiao \href{mailto:yqiao@princeton.edu}{yqiao@princeton.edu}\\

\end{centering}

\section{Introduction}

A binaural room impulse response (BRIR) describes characteristics of acoustic wave interactions from a sound source in a room to the torso, head, and ears of a listener. The use of BRIRs has been ubiquitous in many audio applications. For example, in spatial audio reproduction with headphones, BRIRs are used as audio filters to simulate or reproduce an immersive and perceptually plausible sounding environment; in loudspeaker-based applications, the frequency-domain counterparts of BRIRs are equivalent to the acoustic transfer functions between the loudspeakers and the listener's ears, based on which audio filters are designed for tasks such as crosstalk cancellation (\cite{cooper1989prospects},\cite{gardner19983},\cite{choueiri2018XTC}), room correction/loudspeaker equalization (\cite{karjalainen1999comparison},\cite{lindfors2022loudspeaker}), and personal sound zones (\cite{druyvesteyn1997personal},\cite{betlehem2015personal},\cite{qiao2023effects}). In addition to audio reproduction and rendering, BRIRs have also played an important role in other audio-related tasks, such as sound source localization \citep{shinn2005localizing}, sound source separation \citep{yu2016localization}, and audio-visual learning \citep{younes2023catch}.

As implied by its name, a BRIR is dependent on both the listener's anthropometric features (e.g., ear size and shape) and the room's geometry and acoustic properties. Due to the complex acoustic interactions, such as sound reflections in the room and scattering off the listener, a BRIR varies with both the listener's position and orientation in the room. This is unlike room impulse response (RIR) which only depends on the position, or anechoic head-related impulse response (HRIR) which, in the far-field case, only depends on the orientation. Although there have been multiple HRIR and RIR datasets (\cite{sridhar2017database}, \cite{brinkmann2019cross}, \cite{koyama2021meshrir}) available, and it is possible to synthesize BRIRs from HRIRs and RIRs using methods such as the image source model \citep{wendt2014computationally}, synthesized BRIRs lose physical accuracy and can only maintain perceptual plausibility. While this is sufficient for some applications, such as headphone-based auralization, it is not appropriate for others, such as crosstalk cancellation and personal sound zones, where measured BRIRs are required. Moreover, the lack of a high-resolution BRIR dataset, compared to the existing ones measured at sparse listener positions and orientations (\cite{jeub2009binaural}, \cite{kayser2009database}, \cite{erbes2015database}), has limited the study of BRIR modeling and interpolation at high frequencies and the development of machine learning-based audio applications.

In this paper, we introduce a dataset that contains BRIRs measured in an acoustically treated listening room using multiple loudspeakers and at high-resolution translational and rotational head coordinates. Although the dataset is only measured in one specific room, it is expected to be useful for a wide range of applications as it faithfully captures the spatial dependency of BRIRs on listener positions and orientations. 
For example, it can be directly applied to studies of BRIR modeling and interpolation and applications that require multi-loudspeaker BRIRs, such as crosstalk cancellation and personal sound zones. It has been shown \citep{qiao2023optimal} that the spatial sampling resolution of the dataset is adequate for rendering personal sound zones with continuous listener movements within a certain frequency. With proper interpolation between the measured BRIRs, the dataset can also be used to simulate binaural audio for headphone-based auralization with continuous listener movements. In addition, the dataset can be used for either data augmentation or performance evaluation in a wide range of machine learning-based tasks that require binaural audio with listener movements in multiple degrees of freedom.

\section{Methods}

\subsection{Data Collection}
We measured BRIRs in an irregular-shaped listening room with a near-shoebox shape (see Fig.~\ref{fig:system_setup} for the exact dimensions). The listening room has a $RT_{60}$ of 0.24 s averaged in the range between 1300 and 6300 Hz. The room floor is covered with carpet, and the walls and ceiling are partially covered with acoustic panels. Fig.~\ref{fig:system_setup} also shows the setup and dimensions of the measurement system. A linear array of eight loudspeakers was used as sound sources, and each is a Focal Shape 40 4-inch Flax woofer. The loudspeaker array layout was initially intended for sound field control applications, such as rendering personal sound zones. A Brüel \& Kjær Head and Torso Simulator (HATS, Type 4100) was used as the mannequin listener, with its built-in microphones replaced with a pair of in-ear binaural microphones (Theoretica Applied Physics BACCH-BM Pro). The microphones were calibrated and free-field equalized before the measurement. A custom-made, computer-controlled mechanical translation platform was applied to enable translational movements, and a turntable (Outline ET250-3D) was mounted on top of the platform for rotational movements in the azimuth. 

The BRIR measurement grid has a range of [0.5, 1.0] m in the $y$ direction (front/back) and [-0.5, 0.5] m in the $x$ direction (left/right), with a 0.05 m spacing between adjacent grid points. The distances are relative to the center of the loudspeaker array. At each grid point, the BRIRs were measured at 37 different azimuth angles from the listener facing left to facing right, with a 5$^\circ$ spacing between adjacent angles. In total, there are 68376 (= 11 $y$-translations $\times$ 21 $x$-translations $\times$ 37 azimuthal rotations $\times$ 8 loudspeakers) BRIRs measured.

We measured BRIRs by playing back exponential sine sweep (ESS) signals from the loudspeakers and recording the signals received with binaural microphones. Each sine sweep signal has a length of 500 ms at a 48 kHz sampling rate and is generated using the synchronized ESS method \citep{novak2015synchronized}, with a start frequency of 100 Hz and an end frequency of 24 kHz. The synchronized ESS method is a variant of the traditional ESS method \citep{farina2000simultaneous}, with the advantage of correctly estimating higher harmonic frequency responses. All eight loudspeakers were triggered in series with no overlapping between the ESS signals.

The entire data collection process was split into multiple measurement sessions. For each session, we manually fixed the distance from the listener to the array in the $y$ direction and automated the movements in the $x$ direction and the azimuthal rotations. The measurement automation, signal generation, and data collection were implemented in Cycling '74 Max 8. The BRIR post-processing was performed in MATLAB. Each session lasted for approximately 2 hours, and the entire data collection process took 9 days.

\subsection{Data Processing}

The BRIRs were obtained by first deconvolving the recorded signals with the ESS signal in the frequency domain, with 32768-length FFT at a 48 kHz sampling rate. Then, a fourth-order highpass Butterworth filter with a cutoff frequency of 100 Hz was applied to the deconvolved signals to remove the low-frequency noise that was present during the measurement. Finally, the deconvolved signals were truncated to the first 16384 samples (corresponding to 341.3 ms) and globally normalized. No loudspeaker equalization was applied to the BRIRs as the loudspeaker-specific information, such as directivity, is an integral part of the BRIR and therefore is difficult to compensate for. The processed BRIRs, together with the corresponding listener position and orientation coordinates, were saved as separate files corresponding to different $y$ translations in the SOFA (spatially-oriented format for acoustics, \cite{AES69-2022}) format, following the AES69-2022 (SOFA 2.1) Standard. The dataset was generated using SOFA Toolbox for MATLAB/Octave version 2.2.0.

\section{Data Visualization}

\begin{figure}[h]
    \centering
    \includegraphics[width=0.9\textwidth]{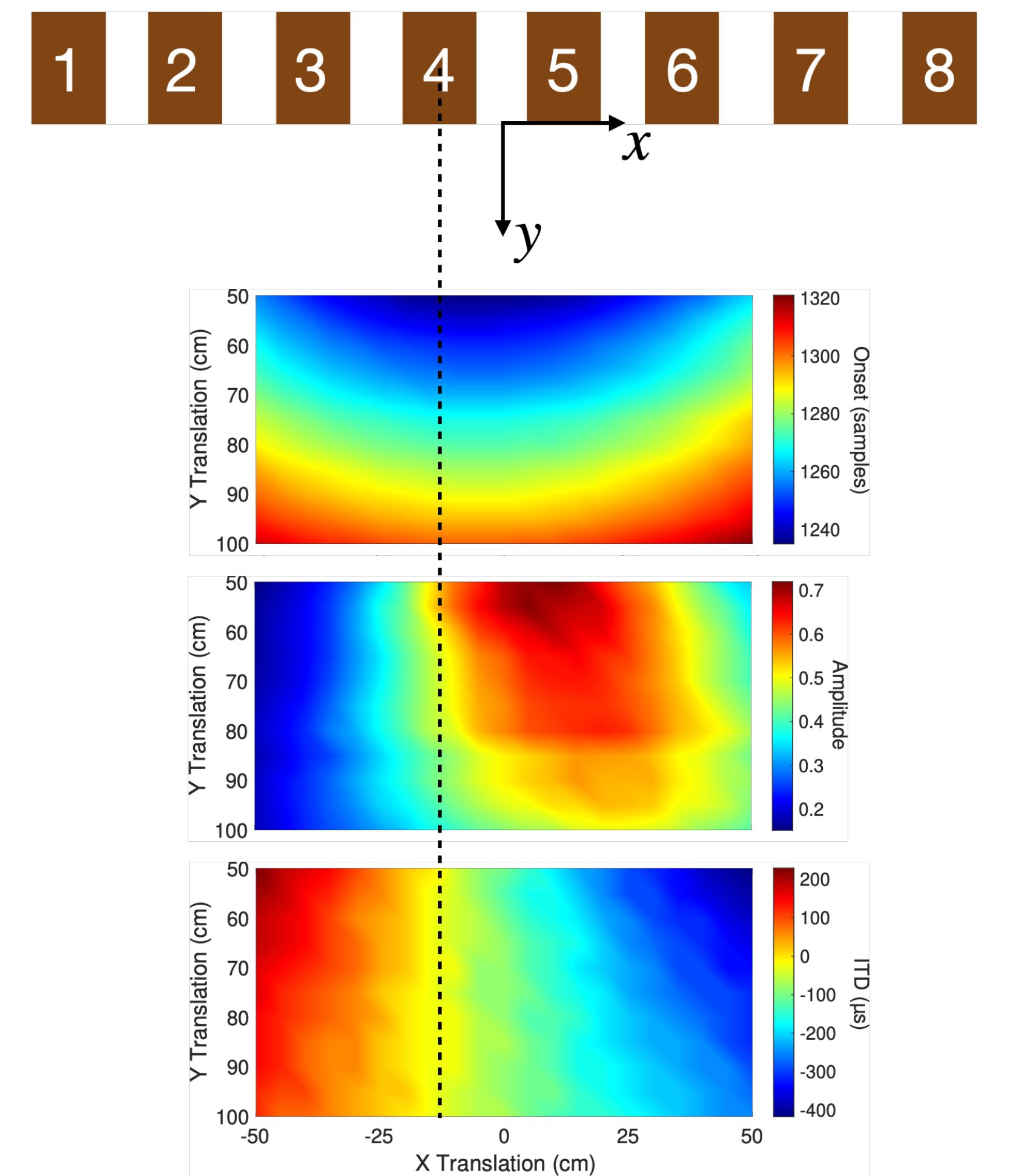}
    \caption{Three example figures showing the properties of the BRIRs measured at the fourth loudspeaker (counting from the left) and with the listener facing forward. The $x$ axis of the figures is aligned with the loudspeaker array. The dashed line in the figure indicates the center of the fourth loudspeaker. Top figure: Onsets of the left-ear BRIRs. Middle figure: Peak amplitude of the left-ear BRIRs. Bottom figure: Calculated ITD of the BRIRs.}
    \label{fig:example_plots}
\end{figure}

We examine the dataset by visualizing 1) the time index of the BRIR onset, 2) the peak amplitude of the BRIR, and 3) the interaural time difference (ITD) of the BRIR, as functions of the listener position. Both the onset and the ITD of BRIRs were calculated in a thresholding approach (see \cite{katz2014comparative} as an example). Fig.~\ref{fig:example_plots} shows the onset, peak amplitude, and ITD of the BRIRs measured at the fourth loudspeaker (counting from the left) and with the listener facing forward. The colors in the figures were interpolated using the ``interp'' option of the MATLAB ``pcolor'' function. All three figures show clear spatial dependency on the listener's position, with the onset increasing and the peak amplitude decreasing as the listener moves away from the loudspeaker. The ITD is nearly zero when the listener is on-axis with the loudspeaker and increases as the listener moves to off-axis positions. Note that the maxima of the peak amplitude in Fig.~\ref{fig:example_plots} are not on-axis with the loudspeaker, which is due to the occlusion effect of the listener's head and ear.

\section*{Data Availability Statement}
The dataset is publicly available at \url{https://doi.org/10.34770/6gc9-5787}.

\section*{Funding}
This work was supported by a research grant from Masimo Corporation.

\section*{Acknowledgments}
We wish to thank K. Tworek, R. Sridhar, and J. Tylka for their prior work on developing the control software for the translation platform and the turntable. We also wish to thank L. Guadagnin for his support on the loudspeaker array.

\bibliographystyle{unsrtnat}
\bibliography{refs}


\begin{figure}[htbp]
    \centering
    \begin{subfigure}[b]{0.6\textwidth}
        \centering
        \includegraphics[width=\textwidth]{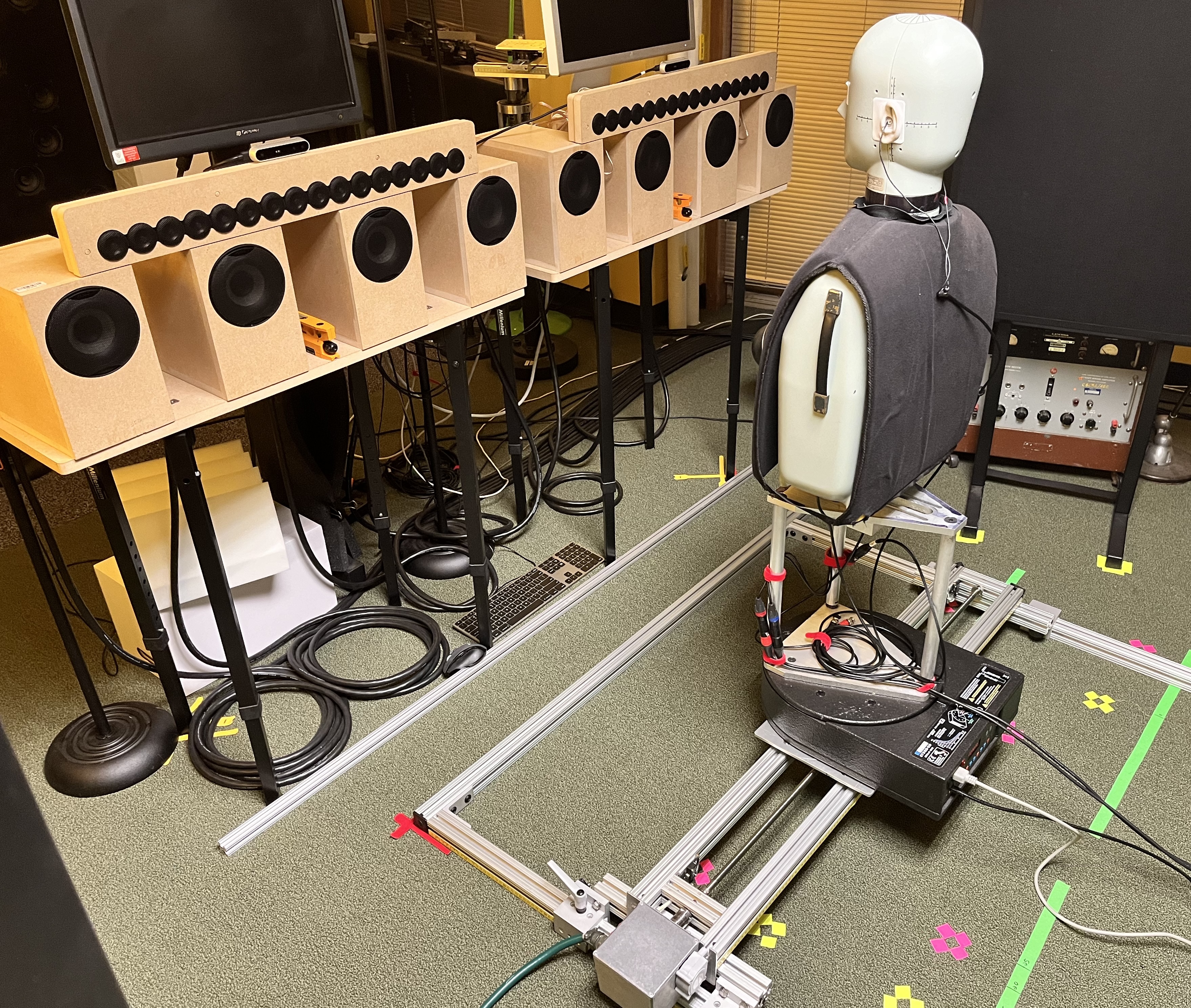}
        \caption{Photo of the measurement system. Note that the tweeter loudspeaker arrays in the photo were not used in the data collection.}
    \end{subfigure}
    \par\bigskip\bigskip
    \begin{subfigure}[b]{0.7\textwidth}
        \centering
        \includegraphics[width=\textwidth]{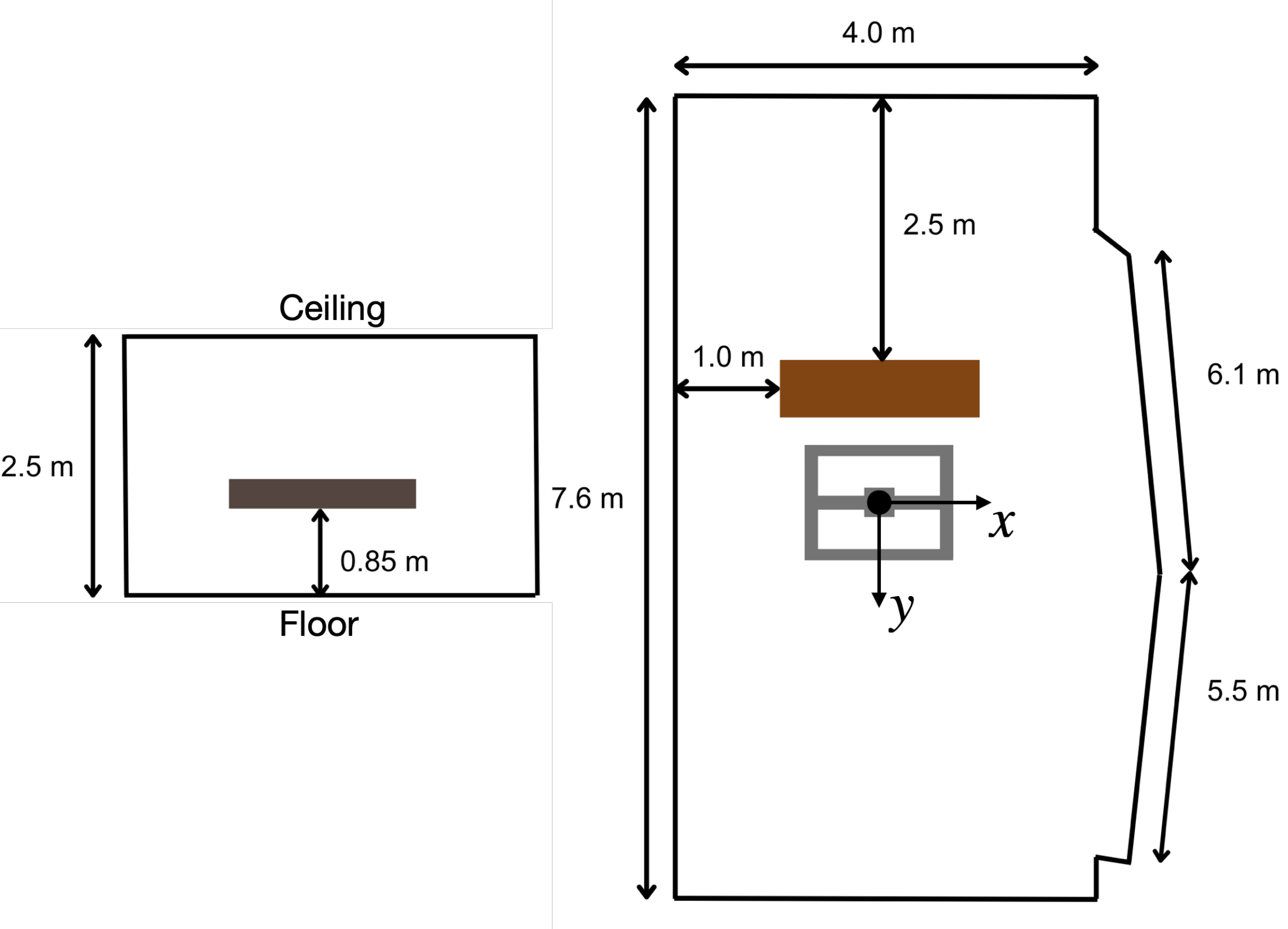}
        \caption{Schematic of the listening room. Left: side view. Right: top view.}
    \end{subfigure}
\end{figure}
    
\begin{figure}[htbp]\ContinuedFloat
    \centering
    \begin{subfigure}[b]{0.7\textwidth}
        \centering
        \includegraphics[width=\textwidth]{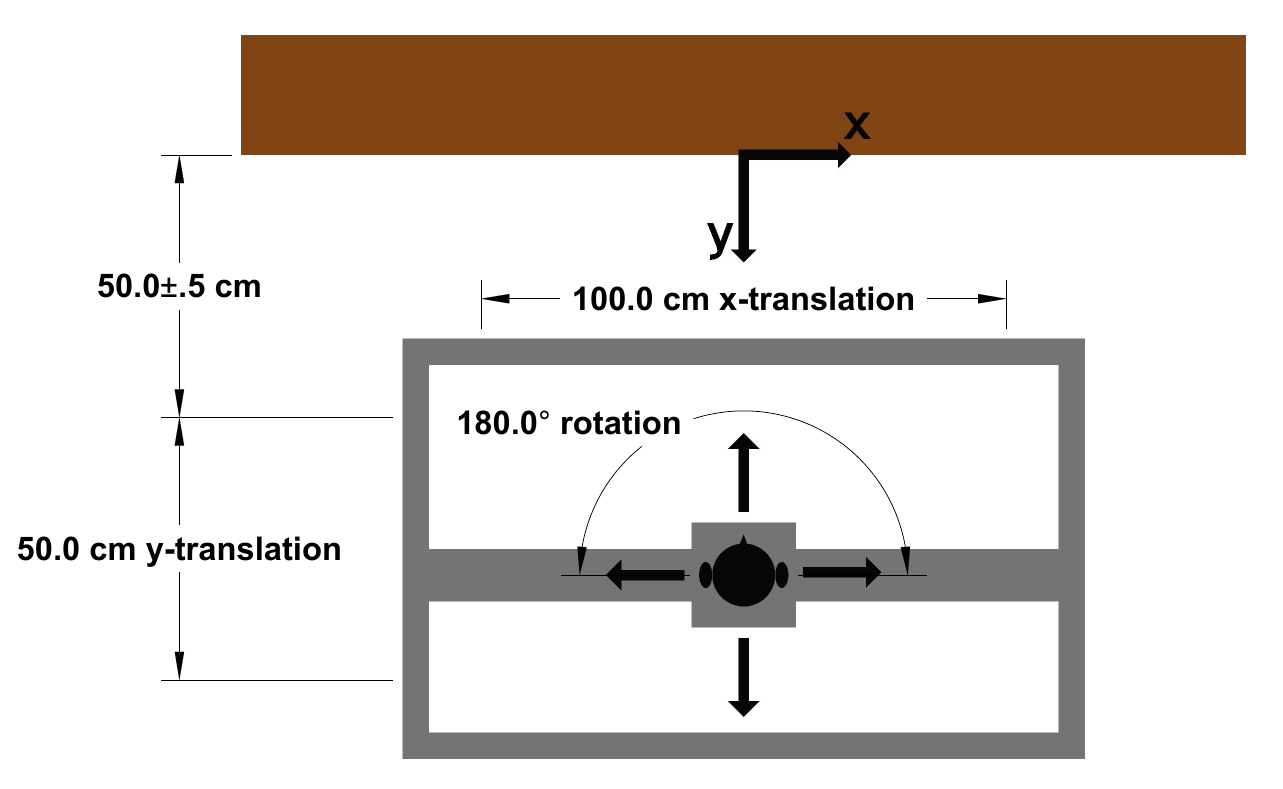}
        \caption{Schematic diagram of the measurement system.}
    \end{subfigure}
    \par\bigskip\bigskip\bigskip
    \begin{subfigure}[b]{0.7\textwidth}
        \centering
        \includegraphics[width=\textwidth]{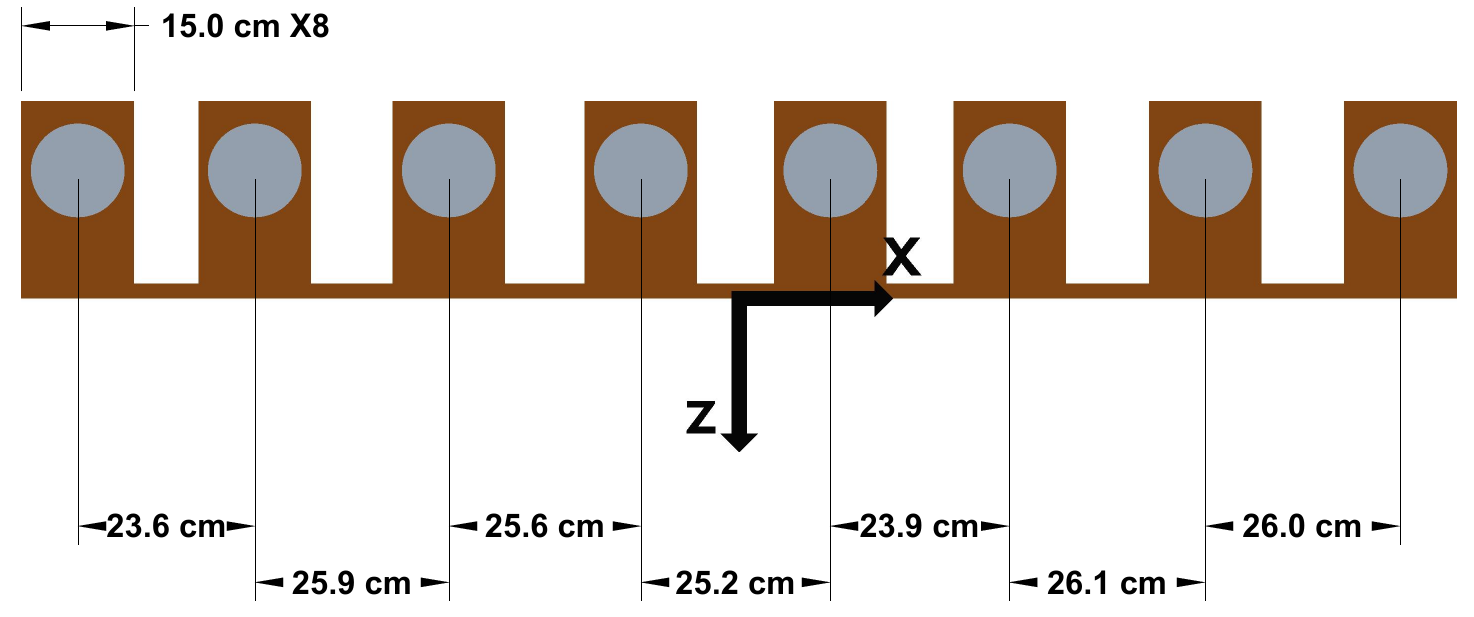}
        \caption{Dimensions of the loudspeaker array.}
    \end{subfigure}
\caption{Illustrations of the measurement system.}\label{fig:system_setup}
\end{figure}

\end{document}